\documentclass[12pt, final]{l4dc2023}

\title[Multi-Agent Reinforcement Learning for Distributed Event-Triggered Control]{Toward Multi-Agent Reinforcement Learning\\ for Distributed Event-Triggered Control}
\usepackage[english]{babel}
\usepackage[utf8]{inputenc}
\usepackage{balance}
\usepackage{xspace}
\usepackage{bbm}
\usepackage{rotating}
\usepackage{url}
\usepackage{csquotes}
\usepackage{paralist}
\usepackage{multirow}
\usepackage{multicol}
\usepackage{amsmath,amssymb,mathtools}
\usepackage{booktabs}
\usepackage{xfrac}
\usepackage{array}
\usepackage{algorithm}
\usepackage[noend]{algorithmic}
\usepackage{textcomp}
\usepackage{siunitx}
\usepackage{microtype}
\usepackage{pgfplots}
\pgfplotsset{compat=newest,unit code/.code={\si{#1}},plot coordinates/math parser=false,grid style={lightgray}, ylabel right/.style={
        after end axis/.append code={
            \node [rotate=90, anchor=north] at (rel axis cs:1,0.5) {#1};
        }   
    }}
\usepgfplotslibrary{units,external,groupplots,fillbetween}
\usetikzlibrary{positioning,angles,quotes,patterns,shapes,spy, shapes.misc,backgrounds}
\usetikzlibrary{external}

\usepackage{scalerel,stackengine}

\newcommand{\fakepar}[1]{\vspace{1mm}\noindent\textbf{#1.}}

\DeclareSIUnit{\belmilliwatt}{Bm}
\DeclareSIUnit{\dBm}{\deci\belmilliwatt}

\DeclareMathOperator*{\E}{\mathbb{E}}

\DeclareMathOperator*{\R}{\mathbb{R}}

\let\originalleft\left
\let\originalright\right
\renewcommand{\left}{\mathopen{}\mathclose\bgroup\originalleft}
\renewcommand{\right}{\aftergroup\egroup\originalright}

\newcommand\figref[1]{Fig.~\ref{#1}}

\newcommand{\eg}{e.g.,\xspace}
\newcommand{\ie}{i.e.,\xspace}

\newcommand{\capt}[1]{\mdseries{\emph{#1}}}

\usepackage{ifthen}
\newboolean{authnotes}

\setboolean{authnotes}{true}

\ifthenelse{\boolean{authnotes}}
{

\newcommand{\db}[1]{\footnote{{\bf\color{green!50!black} Dominik: #1}}}
\newcommand{\st}[1]{\footnote{{\bf\color{purple!90!black} Sebastian: #1}}}
}
{
\newcommand{\am}[1]{}
\newcommand{\db}[1]{}
\newcommand{\st}[1]{}
\newcommand{\mt}[1]{}
}

\newlength\myindent
\usepackage{graphbox}
\usepackage{times}
\usepackage{multicol}
\author{%
 {\Name{Lukas Kesper} \Email{lukas.kesper@rwth-aachen.de}\\
  \Name{Sebastian Trimpe} \Email{trimpe@dsme.rwth-aachen.de}\\
  \addr Institute for Data Science in Mechanical Engineering, \newline
    RWTH Aachen University, Germany}
 \AND
 \Name{Dominik Baumann} \Email{dominik.baumann@aalto.fi}\\
 \addr Department of Electrical Engineering and Automation, \newline 
 Aalto University, Espoo, Finland, \newline
 Department of Information Technology, \newline
 Uppsala University, Sweden%
}

\usepackage{fancyhdr}
\newcommand{\mytitle}{\textbf{Accepted final version.}
To appear in \textit{Proc.\ of the Conference on Learning for Dynamics and Control, 2023}.\\
\copyright 2023 L.~Kesper, S.~Trimpe, and D.~Baumann.}
\begin{document}

\maketitle


\begin{abstract}
Event-triggered communication and control provide high control performance in networked control systems without overloading the communication network.
However, most approaches require precise mathematical models of the system dynamics, which may not always be available.
Model-free learning of communication and control policies provides an alternative. Nevertheless, existing methods typically consider single-agent settings.
This paper proposes a model-free reinforcement learning algorithm that jointly learns resource-aware communication and control policies for distributed multi-agent systems from data.
We evaluate the algorithm in a high-dimensional and nonlinear simulation example and discuss promising avenues for further research.
\end{abstract}


\thispagestyle{fancy}   
\pagestyle{empty}

\section{Introduction}

In networked multi-agent systems, frequent communication of all agents can overload the network, resulting in longer delays and increased risk of message loss \citep{hespanha2007survey}. 
In response, event-triggered methods have been developed \citep{heemels2012introduction,grune14event,miskowicz15event}. 
In event-triggered control (ETC), agents only communicate in case they have something relevant to say. 
Designs for when agents should communicate and how information from other agents should be incorporated into the individual agent's control law are typically based on accurate dynamics models. 
However, such models may not be readily available for high-dimensional and nonlinear, e.g., robotic, systems. 
Moreover, even if they were, jointly optimizing both communication and control strategies may be intractable for complex systems.
Nevertheless, as the separation principle does not necessarily hold for ETC \citep{ramesh2011dual}, such a joint optimization is required to obtain an optimal overall strategy.
Both challenges can be addressed by learning joint communication and control strategies from data through reinforcement learning (RL).

Over the last years, the first algorithms have been developed that learn communication and control strategies from data, see, for instance, \cite{Vamvoudakis.2018,yang2017event,zhong2014event}.
However, most approaches simplify the problem by, \eg imposing a specific structure for the communication strategy, thereby limiting the possibility of the RL algorithm to learn a jointly optimal policy.
A key challenge in learning resource-aware communication and control strategies as compared to learning controllers under periodic communication is the hybrid action space.
At each time step, an agent has to make a discrete communication decision, whether or not to communicate, and what control input to apply, which is typically a continuous value.
Standard RL approaches struggle with this hybrid action space.
A possible remedy is hierarchical RL \citep{sutton1999between}.
The ability of hierarchical RL algorithms to learn high-performing communication and control policies for high-dimensional and nonlinear systems has been demonstrated by \cite{funk2021learning}.
Nevertheless, the authors consider a single-agent system: one agent that is connected to a remote controller over a network.
While this is undoubtedly a relevant problem setting, the issue of resource savings becomes even more paramount when considering distributed multi-agent systems.

Multi-agent RL (MARL), even without communication constraints, is a challenging research field of its own, \cite{zhang2021multi} provide a general overview.
Including resource-aware communication as a design goal adds another complexity layer.
Thus, only few approaches exist that consider learning resource-aware communication and control strategies for multi-agent systems.
The few existing approaches either separate the problem, \ie learn individual communication and control policy \citep{Demirel.2018,lima2022model} or have only been evaluated in low-dimensional systems \citep{Shibata.2021,shibata2022deep}.

\fakepar{Contributions}
We tackle the problem of jointly learning communication and control strategies for high-dimensional networked multi-agent systems. 
In particular, we leverage techniques from hierarchical RL and MARL to learn resource-aware communication and control strategies and
demonstrate the effectiveness of the resulting algorithm in a high-dimensional and nonlinear simulation example.


\section{Related Work}

In the following, we relate our contributions to the literature.

\fakepar{Multi-agent reinforcement learning}
Learning control policies from data is a rapidly evolving research area. 
Various algorithms exist that can learn high-performing control policies without any prior knowledge about the system model \citep{Lillicrap.10.09.2015,Duan.22.04.2016}.
However, the majority of works consider single-agent systems.
Also in MARL, recent works have shown promising results \citep{zhang2021multi, Yu.2020, Tang.25.09.2018,Shu.29.09.2018,Gupta.2017,Lowe.2017}. 
Nevertheless, all mentioned works, both for single-agent and for multi-agent systems, assume that information can be exchanged at high periodic rates.
Such frequent communication may be unsustainable in networked systems where information needs to be transmitted over communication networks.
Thus, herein, we propose an algorithm that learns resource-aware communication and control policies for multi-agent systems.

\fakepar{Learning ETC for single agent-systems} 
For an introduction into and an overview of ETC, we refer the reader to \cite{heemels2012introduction,miskowicz15event}.
In all works discussed therein, the policy design is based on a system model that is assumed to be known.
When this assumption is violated, learning approaches provide an alternative.
Learning ETC from data has received increasing attention in recent years \citep{Sedghi.2022}. 
Many existing approaches constrain the problem, for instance, by predefining the structure of the event-trigger \citep{Vamvoudakis.2018,zhong2014event,yang2017event,Sahoo.2016}.
Restricting the space of communication policies may lead to suboptimal policies, mainly because the separation principle does not generally hold in ETC \citep{ramesh2011dual}.
That is, for obtaining optimal communication and control policies, both need to be optimized jointly.
Such a joint optimization is, \eg done by \cite{baumann2018deep, Hashimoto.2021,funk2021learning}. 
However, those works, as the ones we discussed before, consider single-agent systems. 
Herein, we seek to go a step further and target multi-agent systems.

\fakepar{Learning ETC for multi-agent systems}
Alternative approaches that aim at learning resource-aware control for multi-agent systems are sparse. \cite{Demirel.2018} propose learning a scheduling strategy for a multi-agent control system. 
They use a step-by-step approach in which they begin with designing optimal control policies for each agent, followed by a central agent learning to adapt to the subsystems and controllers. 
A similar approach was chosen by \cite{Hu.2021}, where agents, connected via a network, learn a control policy and then a separate gating policy for the network. 
Both approaches thus separate the optimization problem which might yield suboptimal policies since the separation principle does not hold.
\cite{lima2022model} learn a resource-allocation and control strategy for a multi-agent system assuming both a centralized allocation mechanism and controller, hence, not a distributed setting.
To the best of our knowledge, only one alternative approach exists that jointly optimizes communication and control strategies for distributed multi-agent systems \citep{Shibata.2021,shibata2022deep}. 
Nevertheless, their approach is, in essence, based on the approach by \cite{baumann2018deep}.
Already \cite{baumann2018deep} stated that the approach has troubles scaling to high-dimensional systems.
Consequently, also \cite{Shibata.2021,shibata2022deep} show results only in relatively low-dimensional environments.
In this work, we instead leverage the more scalable approach introduced by \cite{funk2021learning} based on hierarchical RL to design an algorithm that can learn resource-aware communication and control policies for multi-agent systems and seamlessly scales to high-dimensional environments.



\section{Problem Setting}

\begin{figure}
    \centering
    \includegraphics[width=0.4\textwidth, align=c]{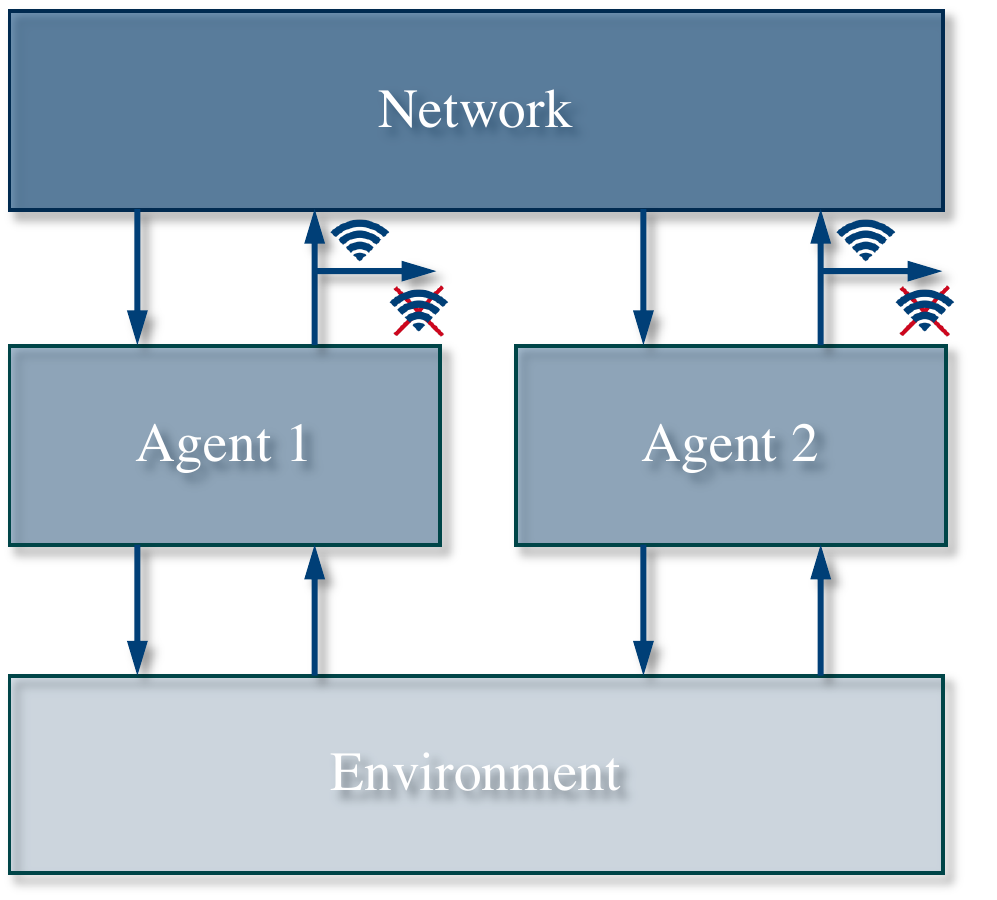}
    \hspace{0.5cm}
    \includegraphics[width=0.4\textwidth, align=c]{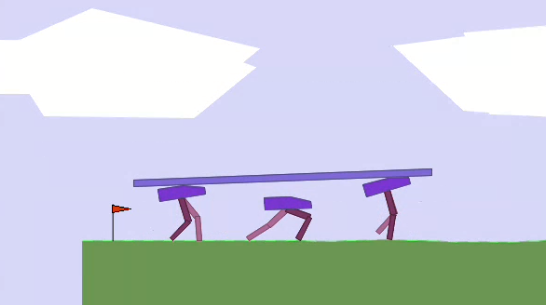}
    \caption{Basic components of the setting and image of the simulation environment. \capt{Left: Agents connected via a network, here for two agents, interacting in an environment. Right: The Multiwalker environment from PettingZoo \citep{Terry.2021} used in our evaluation.}}
    \label{fig:simple_setting}
\end{figure}

We consider a setting with $N$ agents, where the dynamics of each agent $i$ are given by
\begin{equation}
        x_i\left[k+1\right] = f(x\left[k\right], u_i\left[k\right], v_i\left[k\right])
    \label{eq:system_dynamics},
    \end{equation}
with $k\in\mathbb{N}$ the discrete time step, $x_i[k]\in\mathbb{R}^n$ the state of agent $i$, $x[k]\in\mathbb{R}^{N\cdot n}$ the concatenated states of all agents, $u_i[k]\in\mathbb{R}^m$ the action of agent $i$, and $v_i[k]\in\mathbb{R}^n$ process noise.
The agents are supposed to achieve a common control goal, \ie apart from potential couplings through the dynamics function $f$, agents are also coupled through a joint objective.
Thus, while each agent can measure its individual state and select a local control input, they also require information from the other agents for optimal decision-making.
This information is exchanged via a communication network.
The whole setup is shown in Fig.~\ref{fig:simple_setting} (left).

At each time step, agent $i$ needs to decide whether or not to communicate with other agents and which control input to apply, forming a hybrid action space.
Hence, we have two policies: policy $\mu_i$, which decides about sharing information via the network, and policy $\pi_i$, which decides which action $u_i$ to take, \ie which control input to choose. 

\fakepar{Networked communication}
Whenever agents decide to share data in the network, this information is shared among all agents. Thus, the last broadcasted state $\tilde{x}_{i}[k]$ of each agent $i$ is
    \begin{equation}
        \tilde{x}_{i}\left[k\right] =
        \begin{cases}
        x_{i}\left[k\right], & \text{if} \; o_{i} = 1 \\
        \tilde{x}_{i}\left[k-1\right], & \text{if} \; o_{i} = 0,
        \end{cases}
    \label{eq:x_tilde_definition}
    \end{equation}
where $o_i\!\in\!\{0,1\}$ is agent $i$'s communication decision. Implementations of many-to-all communication in wireless control have been demonstrated by \cite{baumann.2021, Mager2019}.

\fakepar{Reward function}
The optimization objective for each agent is formalized through the reward function.
Unlike standard RL designs, we here explicitly add a reward for saving communication.
Thus, we have a standard reward for control performance, $r^\mathrm{ctrl}_i$, rewarding both joint and individual control goals of the agents, and a reward $r^\mathrm{comm}_i=-o_i\xi$, with $\xi\in\R_{\ge0}$, which imposes a cost for communication.
The overall expected discounted return $R_i$ at time step $K$ is then
    \begin{equation}
    R_i[K] = \E\left[\sum\limits_{k=1}^K \gamma^k (r^\mathrm{ctrl}_i\left[k\right] + r_i^\mathrm{comm})\right] = \E\left[\sum\limits_{k=1}^K \gamma^k (r^\mathrm{ctrl}_i\left[k\right]-o_i\left[k\right]\xi)\right]
    \label{eq:Reward_function},
    \end{equation}
where $\gamma\in(0,1)$ is a discount factor.

\fakepar{Problem formulation}
We aim to develop a framework that lets agents maximize the expected discounted return~\eqref{eq:Reward_function} by optimizing policies $\mu_i$ and $\pi_i$.
To achieve optimal performance, we optimize both policies jointly. 
Ultimately, this leads to a framework that can learn high-performing and resource-aware communication and control policies for \emph{(i)} linear and nonlinear, \emph{(ii)} low- and high-dimensional cooperative multi-agent environments with direct interaction between agents, but \emph{(iii)} without the need for a dynamics model.


\section{Background: Hierarchical Reinforcement Learning}
The key to learning event-triggered communication and control policies through RL is handling the hybrid action space, consisting of discrete communication decisions and continuous control tasks. Hierarchical reinforcement learning naturally captures this structure by adding another decision layer above the continuous actions within a Markov decision process \citep{sutton1999between}. This layer consists of discrete options representing primitive decisions with long-lasting effects \citep{Precup.2000}. 
Relating to our problem setting, we have the option to communicate or not to communicate.
Each option is followed by actions from a policy explicitly related to the option. This policy outputs actions until there are better options to choose from, and, therefore, the option is terminated. Trajectory sampling following this structure can be used to train RL agents based on policy gradient theorems provided by the option-critic architecture \citep{Bacon.16.09.2016}. \cite{funk2021learning} use this architecture to derive an algorithm that learns single-agent event-triggered control. Following \cite{funk2021learning}, we now outline the fundamentals of the option-critic architecture.
To keep the notation uncluttered, we consider single-agent systems in this section and, hence, drop the index $i$.

To represent the hierarchy, states $x$ and actions $u$ are extended by options $o$. The architecture uses several policies to parameterize the decision-making, leading from states to the choice of options and actions. Initially, the policy over options $\mu(o\mid x)$ outputs an option, whereupon an associated intra-option policy $\pi_{o}(u\mid x)$ outputs actions. This continues until a termination function $\beta_{o}(x)$ ends the option, leading to another sampling of options by the policy over options. In other words, the policy over options $\mu(o\mid x)$ represents the probability of choosing an option, while the termination function $\beta_{o}(x)$ represents the probability of ending the option. 
The action choice can then be expressed by a single policy,
\begin{equation}
    \pi(u\mid o,x)=(1-\beta_o(x))\pi_{o}(u\mid x)+\beta_o(x)\sum_{\tilde{o}}\mu(\tilde{o}\mid x)\pi_{\tilde{o}}(u\mid x).
\end{equation}
Sampling from this policy yields trajectories from which policy updates can be computed. 
For computing these updates, the expected return~\eqref{eq:Reward_function}, is approximated using a value function. 
In particular, we use a variant of the $Q$-function, which in standard RL methods outputs the expected reward when starting in state $x$, choosing action $u$, and following the according policy from there on. 
In the option critic framework, the $Q$-function also depends on the option, \ie
\begin{equation}
    Q(o,x)=\int\limits_{\tilde{u}}^{}\pi(\tilde{u}\mid x)\Hat{Q}(o,x,\tilde{u})\;\text{d}\tilde{u},
\end{equation}
where $\Hat{Q}(o,x,u)\!=\! r(o,x,u)+\gamma Q(o^{\prime},x^{\prime})$ , with $x^{\prime}$ the successor state of $x$, and $r(o,x,u)$ the reward in a time step. Using the $Q$-function, we can define the return to be maximized through learning, 
\begin{equation}
    R(o,x)=\left[1-\beta_{o}(x)\right]Q(o,x)+\beta_{o}(x)\sum_{\tilde{o}}\mu(\tilde{o}\mid x)Q(\tilde{o}, x ).
\end{equation}
Formulating reward and $Q$-function as functions of state and option allows us to derive gradients to optimize the policies, parametrized by $\theta_\mu$, $\theta_\pi$, and $\theta_\beta$. For the policy over options, taking the gradient of the reward results in 
\begin{equation}
    \frac{\partial R (o,x,\theta_\mu ,\theta_\beta )}
    {\partial\theta_\mu }
    = \beta_{o}(x,\theta_\beta ) \mathbb{E} \left[ \frac{\partial}{\partial \theta_\mu}
    \log (\mu(o,\theta_\mu \mid x)) Q (o,x )\right].
    \label{eq:gradients_mu}
\end{equation}
Besides, it follows from the $Q$-function for the intra-option policy that 
\begin{equation}
    \frac{\partial Q(o,x,\theta_{\pi})}{\partial \theta_\pi}
    = \mathbb{E}\left[\frac{\partial}{\partial \theta_\pi}\log(\pi_{o}(u,\theta_\pi\mid x))\Hat{Q}(o,x,u)\right].
    \label{eq:gradients_pi}
\end{equation}
Following \cite{funk2021learning}, we decide about communication in every single time step. Consequently, we fix $\beta_o(x)\equiv 1$, \ie we terminate options and evaluate the policy over options in every time step. We can now compute policy updates based on the simplified gradients in \eqref{eq:gradients_mu} and \eqref{eq:gradients_pi}. However, we leverage extensions that provide increased performance to improve learning capabilities.

In particular, we use proximal policy option-critic (PPOC) \citep{Klissarov.30.11.2017}, which aims to exploit the performance of proximal policy optimization (PPO) \cite{PPO} in the option-critic architecture. The theory behind PPO is based on trust region policy optimization \citep{TRPO}, where the aim is to maximize a surrogate loss function
\begin{equation}
\label{eq:grad_pi}
    L(\theta)=\mathbb{E}\left[\frac{\pi(u\mid x)}{\pi^\mathrm{old}(u\mid x)}\Hat{A}^{\pi}\right]
\end{equation}
with $\theta$ parametrizing policy $\pi$, and $\Hat{A}^{\pi}$ an estimator of the advantage. This advantage is used to evaluate how beneficial an action is in relation to the expected reward in a time step. PPO extends this loss with a clipping function that clips the value of the ratio 
$\frac{\pi(u\mid x)}{\pi^\mathrm{old}(u\mid x)}$, with a clipping parameter~$\epsilon$. Using the resulting loss function for updates of the intra-option policy, we obtain
\begin{equation} 
    \frac{\partial L(\theta_{\pi})}{\partial(\theta_{\pi})} 
    = \mathbb{E}\left[ \frac{\partial}{\partial\theta_{\pi}} 
    \min[\frac{\pi_{o}(u\vert x)}{\pi^\mathrm{old}_{o}(u\vert x)}A^{\pi}(o,x,u); 
    \text{clip}(\frac{\pi_{o}(u\vert x)}{\pi^\mathrm{old}_{o}(u\vert x)}, 1-\epsilon,1+\epsilon)A^{\pi}(o,x,u)]\right].
    \label{eq:loss_pi}
\end{equation}
Finally, PPOC combines the clipped function~\eqref{eq:loss_pi} with generalized advantage estimation (GAE) \citep{Schulman.2018} for computing the advantage $A^{\pi}(x,o,u)$ to obtain its policy updates. 

\begin{remark}
Generally, also other RL approaches that can handle hybrid action spaces, such as by \cite{neunert2020continuousdiscrete}, could be used. Such approaches have not been extended to ETC yet.
\end{remark}

\section{Learning Resource-Aware Communication and Control for Multi-Agent Systems}

We now present our algorithm.
As discussed previously, we link hierarchical RL and event-triggered communication and control by letting agents share information over a network only if the corresponding option is triggered.
However, the information an agent sends does not directly impact its decision about which control input to apply.
Thus, we omit implementing intra-option policies that are unique to the option chosen by the policy over options $\mu_i$. 
That is, each agent has, independent of the current option, only one control policy $\pi_i$.
Nevertheless, we preserve the general nature of hierarchical RL, since the option choice does have a long-term impact by influencing the behavior of other agents.

In the case of communication, agents share information with all others.
Thus, the available information for each agent consists of its own state $x_i$, as well as the last broadcasted states of the other agents, resulting in an extended state $\hat{x}_{i} = \{x_i, \tilde{x}_1, \dots, \tilde{x}_\mathrm{N}\}$.
An update $\tilde{x}_i$ does not necessarily include all information an agent can send, allowing to send only a subset of information from $x_i$ depending on the environment. The amount of information sent thus becomes an optimization parameter that is relevant for both computing resources and control performance. Nevertheless, receiving state information alone is insufficient for the agents to sample meaningful actions. Agents can only process this information if they are aware of the age of information, \ie how old the last received update is. Thus, we introduce timers, with $\tau_i$ the timer for information sent by agent $i$,
\begin{equation}
    \tau_i\left[k\right]=
        \begin{cases}
        0, & \text{if } o_i[k]=1, \\
        \tau_i\left[k-1\right]+1, & \text{else}.
        \end{cases}
        \label{eq:timer}
\end{equation}
Appending timers $\tau_i$ to last broadcasted states $\tilde{x}_i$ leads to policies $\mu_{i}(o_{i}\mid\hat{x}_{i})$ and $\pi_{i}(u_{i}\mid\hat{x}_{i})$ that enable agents to act within the setting shown in Fig.~\ref{fig:simple_setting}. 

Communication might not have an immediate impact on other agents, which potentially destabilizes learning the policy over options. Thus, we also use PPO to update the policy over options, compared to PPOC, which uses native policy gradients. PPO's sample efficiency can help agents find a resource-aware communication strategy. Utilizing the derivation by \cite{funk2021learning}, we have 
\begin{equation}
    \frac{\partial L(\theta_{\mu_i})}{\partial(\theta_{\mu_i})} 
   \! =\! \frac{\partial}{\partial \theta_{\mu_i}}\mathbb{E}[ 
    \min[\frac{\mu_{i}(o_i\vert \hat{x}_i)}{\mu^\mathrm{old}_{i}(o_i\vert \hat{x}_i)}A^{\mu}_{i}(o_i, \hat{x}_i, x); 
    \text{clip}(\frac{\mu_{i}(o_i\vert \hat{x}_i)}{\mu^\mathrm{old}_{i}(o_i\vert \hat{x}_i)}, 1-\epsilon,1+\epsilon) A^{\mu}_{i}(o_i, \hat{x}_i, x)]],
    \label{eq:loss_mu}
\end{equation} 
where we use a greedy advantage function
\begin{equation}
    A^{\mu}_{i}(o_i, \hat{x}_i, x)=Q_i (o_i, \hat{x}_i, x)-\max_{\tilde{o}_i\in\{0,1\}}Q_i(\tilde{o}_i, \hat{x}_i, x).
    \label{eq:greedy_adv}
\end{equation}
Besides performance improvements, restricting the updates of the policy over options through the clipping function is also a reasonable measure to limit drastic changes in the policy that could potentially destabilize the learning of other agents.

In~\eqref{eq:greedy_adv}, we see that $Q_i$ depends on the concatenated state $x$ in addition to the extended state $\hat{x}_i$.
This is connected to the destabilizing effect of other agents who change their policy through learning.
From the individual agent's perspective, which models other agents as part of the environment, these changes make the environment appear unstable.
In our case, this is an even greater challenge as each agent has several policies. To compensate for this perception, we use centralized learning and decentralized execution (CLDE): during learning, we use a critic for each agent $i$, \ie the $Q_i$-function, to compute policy updates. During execution, the policies $\mu_i$ and $\pi_i$ are used to sample trajectories without the critic. By centralizing the critic, learning about the value function can be improved, helping to stabilize the overall learning process \citep{Lyu.b}. Such centralization can be achieved by handing over all relevant state information from other agents $j$ to the critic of an agent $i$, following the approach by \cite{Lowe.2017}. We implement this by always passing the true states $x_j$ to the critic of agent $i$ in addition to the last broadcasted states $\tilde{x}_j$ to compensate for the lack of information through saving communication.
\begin{remark}
CLDE requires all agents to communicate periodically during training.
We assume that during training in a laboratory environment, we can provide sufficient bandwidth for periodic communication, but during execution in the real world need to save resources.
\end{remark}

We use deep neural networks (DNNs) to approximate the $Q_i$-functions, the policy over options $\mu_i$, and each agent's control policy $\pi_i$. 
We follow the DNN structure proposed by \cite{funk2021learning}.
However, we set the standard deviation of the stochastic control policy to a constant value instead of slowly decreasing it. Lowering the stochasticity aims to reduce the dispersion in the behavior of other agents, making it easier for the agents to adapt to each other. 

This reduced stochasticity also causes less exploration. Thus, we always consider the learning progress over training epochs and introduce a curriculum that supports the agents in generating a solution. First, we apply noise to the actions $u_i$. In particular, we use values from a Gaussian distribution with zero mean and a linearly decreasing standard deviation $\sigma$. This way, we achieve adjustability and phasing for agent exploration, effectively making exploration a hyperparameter. However, while the agents are still exploring the observation space to improve the output of the control policy, the communication is already penalized. 
The policy over options typically reacts quickly by reducing communication. This inevitably leads to local optima, as no cooperative action is possible without shared information. Therefore, we leverage entropy regularization, which ensures that communication is learned only after exploration and, thus, forms the next step in the curriculum. In the beginning, we apply the entropy term $\zeta \log(\mu_i (o_i \mid \hat{x}_i))\mu_i (o_i \mid \hat{x}_i)$ to the updates of the policy over options, limiting updates through the entropy regularization coefficient $\zeta$. After sufficiently exploring the control policy, we lower the value of $\zeta$ linearly over epochs, resulting in slowly enabled updates to the policy over options. After reaching the final epoch of the decrease, the coefficient remains at a low value. Typically, this leads to destabilization and decreased rewards since agents receive less information as communication savings increase. As the last step, we, therefore, introduce another episode of noise in the curriculum, whose standard deviation first increases linearly over epochs and then decreases again. This yields further exploration of the policies $\pi_i$ and $\mu_i$, as well as the $Q_i$-function. In summary, the curriculum proceeds in three steps: we start by \emph{(I)} adding noise to the actions with decreasing variance, followed by \emph{(II)} enabling policy updates of the policy over options, and \emph{(III)} finally, we apply noise again to stimulate full exploration of the observation space.
The entire algorithm is summarized in Alg.~\ref{alg:multi-agent}.

\begin{algorithm}
\caption{Learning algorithm for distributed event-triggered control.
}\label{alg:multi-agent}
\begin{multicols}{2}
\begin{algorithmic}[1]
\STATE \textbf{Input:} clipping $\epsilon$, curriculum param.\ $\zeta, \sigma$ \\
\FOR{number of agents}
    \STATE Initialize $Q_{i}(o_i ,\hat{x}_i ,x), \mu_i(o_{i}\vert\hat{x}_i), \pi_i(u_{i}\vert\hat{x}_i)$
\ENDFOR
\FOR{number of epochs}
    \FOR{number of agents}
        \STATE $Q_{i} \gets Q_{i}^\mathrm{old}$, $\mu_i \gets \mu_i^\mathrm{old}$, $\pi_i \gets \pi_i^\mathrm{old}$
    \ENDFOR
    \STATE Sample $(o,x,u)$-transitions using $\mu_i, \pi_i$
    \FOR{number of agents}
        \STATE Calculate $A^{\pi}_{i}(o_i, \hat{x}_i, x, u)$ using GAE
        \FOR{number of optimizer iterations} 
            \FOR{number of options}
                \STATE Sample batch
                \STATE Calculate $A^{\mu}_{i}(o_i,\hat{x}_i,x)$ from \eqref{eq:greedy_adv}
                \STATE Calculate $L^{\mu}_{i}$ from \eqref{eq:loss_mu}
                \STATE Calculate $L^{\pi}_{i}$ from \eqref{eq:loss_pi}
                \STATE Calculate $L^{Q}_{i}$ as \cite{funk2021learning} 
                
                \STATE $\theta_{\mu_{i}}\gets \theta_{\mu_{i}}+\alpha\frac{\partial L^{\mu}_{i}(\theta_{\mu_{i}})}{\partial\theta_{\mu_{i}}}$
                \STATE $\theta_{\pi_{i}}\gets \theta_{\pi_{i}}+\alpha\frac{\partial L^{\pi}_{i}(\theta_{\pi_{i}})}{\partial\theta_{\pi_{i}}}$
                \STATE $\theta_{Q_{i}} \gets \theta_{Q_{i}}-\alpha\frac{\partial L^{Q}_{i}(\theta_{Q_{i}})}{\partial\theta_{Q_{i}}}$
\ENDFOR
\ENDFOR
\ENDFOR

\STATE \textbf{switch} (curriculum phase)
\STATE \hspace{10pt} \textbf{case} I: Update $\sigma$
\STATE \hspace{10pt} \textbf{case} II: Update $\zeta$
\STATE \hspace{10pt} \textbf{case} III: Update $\sigma$

\ENDFOR
\end{algorithmic}
\end{multicols}
\end{algorithm}


\section{Results and Analysis}

We demonstrate the effectiveness of our framework in a simulation example that includes a traceable cooperative task in a high-dimensional multi-agent environment with complex physics. 
For this, we choose the Multiwalker environment from PettingZoo \citep{Terry.2021}, shown in \figref{fig:simple_setting} (right). 
The goal in this environment is for the three agents to cooperatively transport a package forward, without falling or dropping the package.

We slightly modify the environment to fit our problem setting.
In particular, we change the agents' observation space. 
In the native implementation, this information includes all joint angles, speeds, and contact sensors of the individual agent. Besides, agents use LIDAR sensors to scan the environment and obtain information about the position of the agent in front of them. 
Alongside the LIDAR, agents receive state information about their neighbors in the form of relative positions, as well as their relative position towards the package and the package angle.
To make communication necessary, we remove the LIDAR information. Similarly, we remove relative information about neighbors, which we obtain through communication.
The state of an individual agent consists then in their own joint angles, speeds, contact forces, and information about the package, which results in 17 state variables per agent.
We introduce communication by allowing agents to share their absolute position and horizontal and vertical velocity, from which receiving agents can calculate relative positions and velocities.
We thus define the last broadcasted state $\tilde{x_i}$ as the relative distances in position and the relative differences in the velocities of the agents, combined with the corresponding timer $\tau_i$. Ultimately, the agents know their own state, relevant relative positions, as well as information about the relative state of the package, all combined in the networked observation $\hat{x}_i$.  

We further need to decide about inputs to the policies.
As we use CLDE, the $Q_i$-functions receive all information available to the agent and the relative data of the other agents that have not been communicated. The policy $\pi_i$ receives the package measurements and the communicated states of other agents to sample meaningful actions.
For the policy $\mu_i$, experiments have shown that it is most effective to limit the input to the package information. The idea is to initiate communication based on how the package behaves during a training run. For instance, a deflecting package can be a relevant trigger for sending an update. Thus, we add only the current and the package information from the last update to the policy over options $\mu_i$. 

Lastly, we define the reward function.
Different from the original environment, we add a communication penalty.
Further, to encourage agents to find creative solutions, we do not penalize head deflection and extend the possible torque for the joints by a factor of $1.5$. In addition, we introduce a small penalty for speed in the horizontal direction to prevent agents from learning risky gaits that, according to our observations, can lead to the collapse of the entire learning process. 
 
\begin{figure}
    \centering
    \includegraphics[height=0.34\textheight]{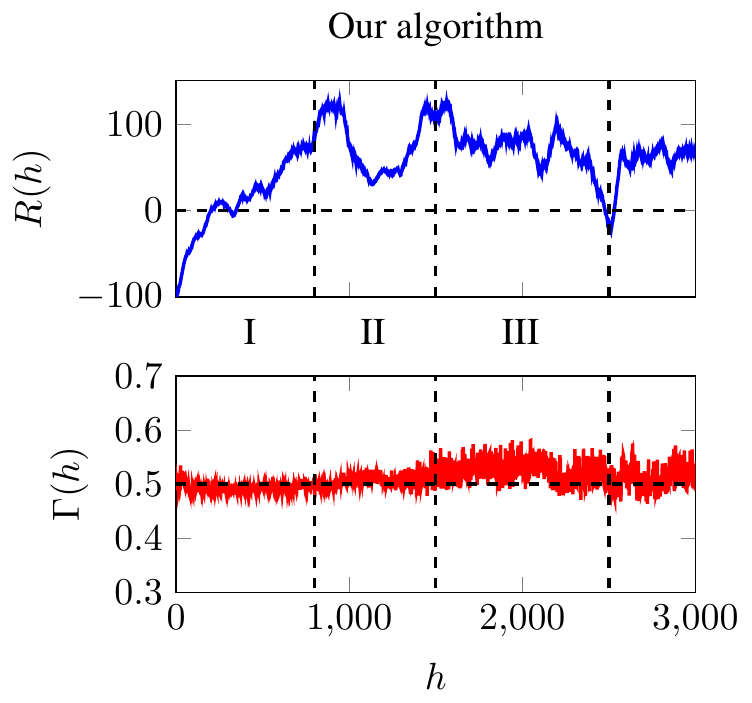}
    \includegraphics[height=0.34\textheight]{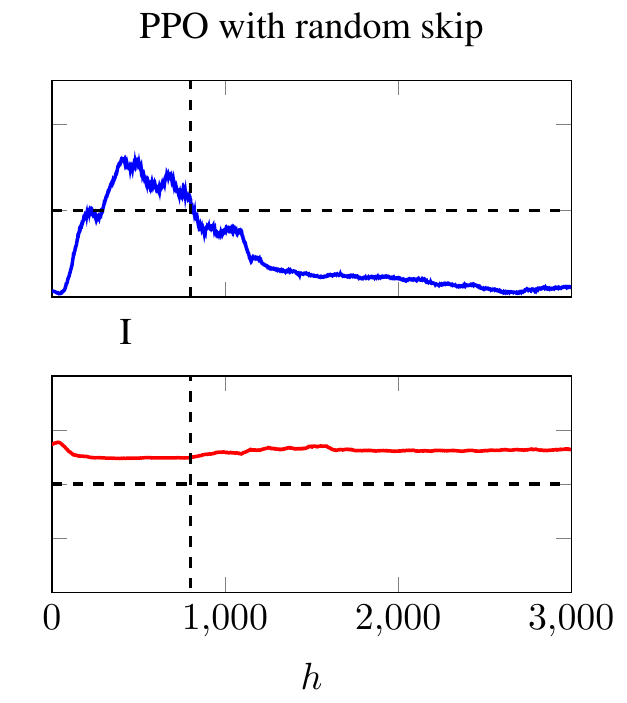}
    \caption{Performance of our algorithm and PPO with random communication skips. \capt{We show the return (top plots) and communication savings (bottom plots). While our algorithm achieves significant communication savings and good performance (left), PPO with random communication skips is unstable and cannot learn a good policy (right). Dashed vertical lines represent the stages of the curriculum.}}
    \label{fig:comparison}
\end{figure}

We train the algorithm for ten runs, where one training iteration includes $2048$ time steps for each agent, and we compute updates over 3000 epochs.
We show the best performing run in Fig.~\ref{fig:comparison} (left).
In particular, we show the return $R(h)$ and communication savings $\Gamma(h)$, calculated in comparison to periodic communication at each time step with an underlying sampling time of \SI{20}{\milli\second}, of all agents over epochs $h$, averaged over 50 epochs. 
We see the effect of the three curriculum stages in the learning curve, marked by dashed vertical lines. 
Initially, we only explore the control policy, leading to increasing reward.
As the agents start to actively reduce communication in the second phase, the reward first drops, but recovers after a while. 
In the third phase, we again explore the control policy, which stabilizes the learning result.
Finally, the agents learn to move the package cooperatively forward while saving around \SI{55}{\percent} of communication.
The resulting policy is intuitive in that, most of the time, only two agents actively work on transporting the package, reducing the need to coordinate\footnote{Videos and code are available at \href{https://sites.google.com/view/learning-distributed-etc/}{https://sites.google.com/view/learning-distributed-etc/}.}.

We further compare our algorithm to native PPO.
\cite{kargar2021macrpo} have shown that PPO, in general, can learn good policies for the Multiwalker.
We thus train PPO on the Multiwalker and randomly skip communication to also achieve savings around \SI{55}{\percent}.
While PPO does not make its own decisions about communication, we leave the remaining design unchanged, \ie we use CLDE, the same DNN structure, and the first stage of the learning curriculum.
We then also execute ten training runs and show the best performing policy in \figref{fig:comparison} (right). 
The return PPO obtains is significantly lower than for our algorithm.
This is particularly remarkable as PPO does not incur any communication penalties.
The PPO agents cannot successfully transport the package forward, demonstrating that the task is non-trivial.


\section{Conclusion}
This article presents a first step toward automatic learning of joint communication and control policies for high-dimensional and distributed multi-agent systems.
We capture the hybrid nature of joint communication and control policies through a hierarchical RL approach and optimize both.
The resulting algorithm significantly outperforms a control policy with random communication skips.

While these are promising results, there are various possibilities for further research.
The algorithm achieves good performance on a challenging simulation task.
However, the training stability may still be improved.
When training multiple agents in parallel, around half of the agents end up with a performance comparable to what we have shown in the previous section.
Through hyperparameter tuning, a higher success rate may still be attainable.
Further, the underlying algorithm and software framework are, in machine learning terms, relatively old.
In particular, we use Tensorflow 1 \citep{abadi2016tensorflow} and the baselines \citep{stable-baselines} implementation of PPO \citep{PPO}.
By updating to more modern frameworks, increased performance and stability might be possible.
The choice for PPO as the underlying RL algorithm was motivated by findings that it performs well in multi-agent settings \citep{yu2021surprising}.
Nevertheless, dedicated MARL algorithms may still provide a further performance increase.
Also, adopting specific MARL measures beyond CLDE, such as parameter sharing among agents \citep{christianos2021scaling}, might yield more stable and successful learning.
Lastly, the options framework allows for even greater flexibility.
For instance, one could imagine options for sending different kinds of information depending on the state of the environment, or combining our approach with event-triggered control updates.


\acks{Special thanks go to Niklas Funk, Friedrich Solowjow, and Bernd Frauenknecht for their technical input, as well as to Jonas Reiher, Johanna Menn, and Johannes Berger for their helpful advice.
This work is funded in part under the Excellence Strategy of the Federal 
Government and the Laender (G:(DE-82)EXS-SF-SFDdM035) of Germany.
Simulations were performed with computing resources granted by RWTH Aachen 
University under project thes1218.}

\bibliography{ref}

\end{document}